# Brightening interlayer excitons by electric-field-driven hole transfer in bilayer WSe$_2$

Tianyi Ouyang[1*], Erfu Liu[1,2*], Soonyoung Cha[1*], Raj Kumar Paudel[3,4], Yiyang Sun[1], Zhaoran Xu[1], Takashi Taniguchi[5], Kenji Watanabe[6], Nathaniel M. Gabor[1†], Yia-Chung Chang[3,4†], and Chun Hung Lui[1†]

[1]Department of Physics and Astronomy, University of California, Riverside, CA 92521, USA
[1]School of Physics, National Laboratory of Solid State Microstructures, and Collaborative Innovation Center for Advanced Microstructures, Nanjing University, Nanjing, 210093, China.
[3]Research Center for Applied Sciences, Academia Sinica, Taipei 11529, Taiwan
[4]Department of Physics, National Cheng-Kung University, Tainan, Taiwan
[5]Research Center for Materials Nanoarchitectonics, National Institute for Materials Science, 1-1 Namiki, Tsukuba 305-0044, Japan
[6]Research Center for Electronic and Optical Materials, National Institute for Materials Science, 1-1 Namiki, Tsukuba 305-0044, Japan

*These authors equally contribute to this work
†Corresponding authors: Corresponding author. Email: nathaniel.gabor@ucr.edu, yiachang@gate.sinica.edu.tw, joshua.lui@ucr.edu

We observe the interlayer $A_{1s}^I$, $A_{2s}^I$, and $B_{1s}^I$ excitons in bilayer WSe$_2$ under applied electric fields using reflectance contrast spectroscopy. Remarkably, these interlayer excitons remain optically bright despite being well separated from symmetry-matched intralayer excitons—a regime where conventional two-level coupling models fail unless unphysically large coupling strengths are assumed. To uncover the origin of this brightening, we perform density functional theory (DFT) calculations and find that the applied electric field distorts the valence-band Bloch states, driving the hole wavefunction from one layer to the other. This field-driven interlayer hole transfer imparts intralayer character to the interlayer excitons, thereby enhancing their oscillator strength without requiring hybridization with bright intralayer states. Simulations confirm that this mechanism accounts for the major contribution to the observed brightness, with excitonic hybridization playing only a minor role. Our results identify interlayer hole transfer as a robust and general mechanism for brightening interlayer excitons in bilayer transition metal dichalcogenides (TMDs), especially when inter- and intra-layer excitons are energetically well separated.

*Introduction*—Two-dimensional (2D) transition metal dichalcogenides (TMDs) exhibit a rich landscape of excitonic phenomena arising from strong Coulomb interactions, reduced dielectric screening, and spin–valley locking [1-7]. In bilayer TMDs, an added spatial degree of freedom enables the formation of interlayer excitons—bound states of electrons and holes residing in different layers. These spatially indirect excitons possess long lifetimes and exhibit field-tunable energies via the Stark effect, making them promising for optoelectronic applications and for exploring dipolar many-body physics [8-14].

Yet, the spatial separation of charge carriers suppresses the oscillator strength of interlayer excitons, limiting their visibility in optical spectroscopy. While they can dominate photoluminescence at low energies due to carrier accumulation, their weak optical coupling renders them nearly invisible in absorption-based probes such as reflectance contrast at higher energies [15-20].

Surprisingly, recent experiments in bilayer $MoS_2$ and $MoSe_2$ have revealed bright interlayer excitons at elevated energies, challenging the conventional expectation of weak optical activity [21-32]. A widely invoked explanation attributes this brightness to hybridization with intralayer excitons, which allows the interlayer states to inherit oscillator strength, as illustrated by Figure 1a [24, 25, 28, 30, 31]. Such models, typically based on constant coupling in two- or multi-level frameworks, offer intuitive descriptions but neglect electric-field-induced modifications of Bloch states, coupling matrix elements, and exciton envelope functions. These simplified models are applicable near resonance, where strong state mixing dominates, but may fail in regimes of large energy detuning where other mechanisms become relevant.

In this Letter, we investigate the exciton spectra of bilayer $WSe_2$ under applied electric fields by reflectance contrast spectroscopy. In contrast to $MoS_2$, bilayer $WSe_2$ exhibits large energy separations: 424 meV between the interlayer $A^I$ and intralayer $B^0$ excitons, and 538 meV between the interlayer $B^I$ and intralayer $A^0$ excitons. These wide separations prevent significant hybridization. Nevertheless, we observe clear optical signatures from the interlayer $A^I_{1s}$ and $B^I_{1s}$ excitons, along with a very weak signal from the $A^I_{2s}$ state, allowing us to extract an interlayer $A^I_{1s}$ binding energy of 92 meV.

To interpret these observations, we evaluate two models. The first is the conventional two-level hybridization model, as illustrated in Fig. 1(a). While it reproduces the observed spectral features, it relies on unrealistically large coupling strengths, rendering it physically implausible. The second model, motivated by *ab initio* calculations and shown in Fig. 1(b), invokes a field-driven interlayer hole-transfer mechanism. Here the applied electric field distorts the valence-band states, gradually shifting the hole wavefunction from the majority to the minority layer. This continuous redistribution imparts intralayer character to the interlayer exciton, thereby enhancing its oscillator strength without requiring hybridization with bright intralayer states.

Our simulations show that this field-driven hole-transfer mechanism accounts for more than 80% of the interlayer $A^I_{1s}$ exciton brightness, with hybridization contributing less than 20% over our experimental field range. These results establish field-driven valence-band distortion as the dominant mechanism for interlayer exciton brightening in bilayer $WSe_2$ in the regime where inter- and intra-layer excitons are energetically well separated. This insight advances our understanding of exciton dynamics in van der Waals materials and provides a framework for engineering bright, long-lived excitons beyond conventional hybridization models.

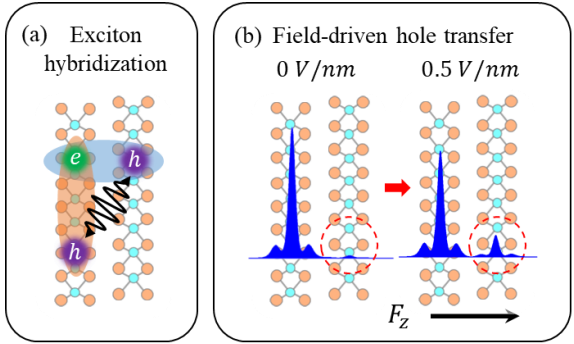

FIG. 1. (a) Schematic illustration of hybridization between intralayer and interlayer excitons in bilayer transition metal dichalcogenides (TMDs) (b) Schematic of field-driven interlayer hole transfer, where an applied electric field distorts the valence-band states and shifts the hole density profile across layers in bilayer TMDs. The hole transfer is highlighted by the dashed circles.

*Experiment*—Our experiments utilize dual-gated bilayer WSe$_2$ devices encapsulated by hexagonal boron nitride (BN) [Fig. 2(a)]. Thin graphite flakes serve as contact and gate electrodes to optimize device performance. By applying voltages of opposite polarity to the top and bottom gates, we generate a vertical electric field across bilayer WSe$_2$ while maintaining overall charge neutrality. Reflection spectra ($R_s$) are measured on the sample area, while a reference spectrum ($R_r$) is taken on a nearby BN-encapsulated area without WSe$_2$. The reflectance contrast is then calculated as $\Delta R/R = (R_s - R_r)/R_r$. To enhance the visibility of weak spectral features, we further compute the second derivative with respect to the photon energy, yielding $d^2(\Delta R/R)/dE^2$ spectra.

Bilayer WSe$_2$ hosts multiple intralayer and interlayer excitons [Fig. 2(b)], which are revealed in the $d^2(\Delta R/R)/dE^2$ map as a function of the screened electric field $F$ [Fig. 2(c)]. The three prominent vertical features correspond to the intralayer $A^0_{1s}$, $A^0_{2s}$, and $B^0_{1s}$ excitons. The $A^0_{1s} - B^0_{1s}$ energy separation is 461 meV, reflecting the strong spin-orbit coupling in WSe$_2$ [2, 6, 33]. At zero electric field, the interlayer excitons are not visible. However, as the field increases, three additional excitonic features emerge—none of which have been reported previously.

The most prominent of these can be traced to 1.732 eV at zero field and exhibits a Stark-induced blueshift, yielding a dipole moment of 0.61 $e·$nm. We identify this feature as the interlayer $A^I_{1s}$ exciton. A second, much weaker line appears 64 meV above $A^I_{1s}$ and exhibits an identical Stark shift, consistent with assignment to the interlayer $A^I_{2s}$ exciton. The third interlayer feature, fainter than $A^I_{1s}$ but stronger than $A^I_{2s}$, is located at 2.241 eV at zero field and displays a redshift under increasing electric field, yielding a dipole moment of 0.6 e·nm. We attribute this line to the interlayer $B^I_{1s}$ exciton.

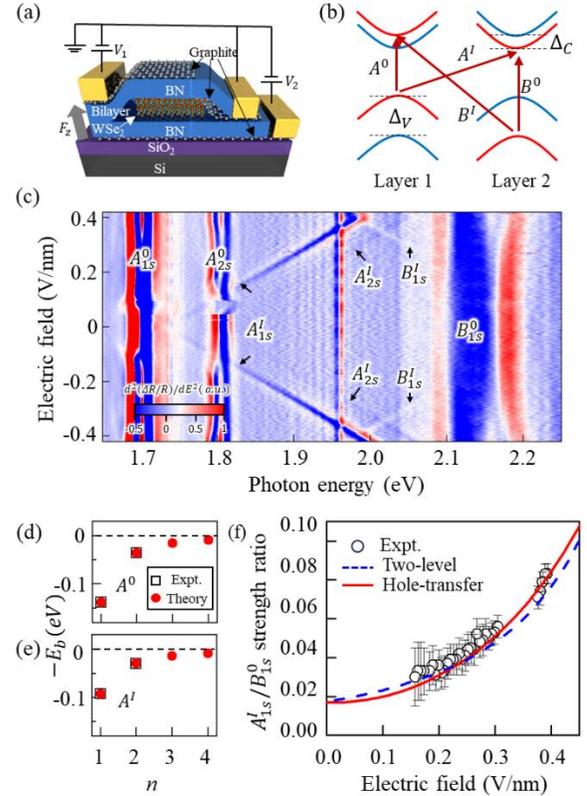

FIG. 2. (a) Schematic of the dual-gated bilayer WSe$_2$ device. (b) Illustration of excitonic band transitions in 2H-bilayer WSe$_2$ (c) Color map of the second derivative of the reflectance contrast with respect to photon energy, plotted as a function of screened electric field across the bilayer. The vertical line at 1.96 eV is an artifact caused by stitching two spectra recorded over different energy ranges. (d-e) DFT-calculated binding energies (red dots) of the 1s-4s Rydberg states of the intralayer (top) and interlayer (bottom) A excitons, obtained by fitting the experimentally measured 1s-2s energy separations (black squares). (f) Oscillator strength of the $A_{1s}^I$ exciton relative to the $B_{1s}^0$ exciton as a function of electric field, extracted from the data in panel (c). The blue and red curves represent simulations based on the two-level coupling model and the interlayer hole transfer model, respectively.

To gain quantitative insight, we combine DFT and empirical Keldysh potentials to fit the Rydberg series for both intra- and inter-layer A excitons (see Supplemental Materials [37] for details). With Keldysh screening lengths parameters $r_0$ = 1.27 nm for intralayer and $r_0$ = 3.06 nm for interlayer excitons, and a common effective dielectric constant κ = 4.4—all within physically reasonable range—we accurately reproduce the experimentally observed $A_{1s}^0 - A_{2s}^0$ energy separation (103 meV) and $A_{1s}^I - A_{2s}^I$ separation (64 meV) [Fig. 2(d)]. The larger screening lengths $r_0$ for $A_{1s}^I$ is expected, as interlayer excitons are more weakly bound and exhibit a crossover from 2D to 3D behavior. From these fits, we extract binding energies of 139 meV and 36 meV for the $A_{1s}^0$ and $A_{2s}^0$ exciton, and 92 meV and 29 meV for the $A_{1s}^I$ and $A_{2s}^I$ excitons, respectively.

Additionally, using the measured generation energies of $A_{1s}^0$, $A_{1s}^I$, $B_{1s}^0$, and $B_{1s}^I$, we extract several key band structure parameters: a free-particle band gap of 1.842 eV, a conduction-band spin-orbit splitting $\Delta_C$ = 17 ± 4 meV, and a valence-band splitting $\Delta_V$ = 458 ± 7 meV [Fig. 2(b)]. These values are consistent with previous reports [34-36] (see Supplemental Materials [37] for details).

*Two-level model*—Next, we turn to the field-dependent behavior of the interlayer excitons. According to the field-dependent intensity of $A_{1s}^I$ shown in Fig. 2(c, f), the $A_{1s}^I$ line becomes bright even when it lies ~300 meV below the $B_{1s}^0$ exciton—a regime well outside the typical level-crossing conditions of exciton hybridization. Previous studies on bilayer MoS$_2$ and MoSe$_2$ have attributed interlayer exciton brightening to hybridization with intralayer excitons via two or multi-level models with constant coupling terms [24, 25, 28, 30, 31]. Such models account for field-dependent behavior in the regime of level crossing.

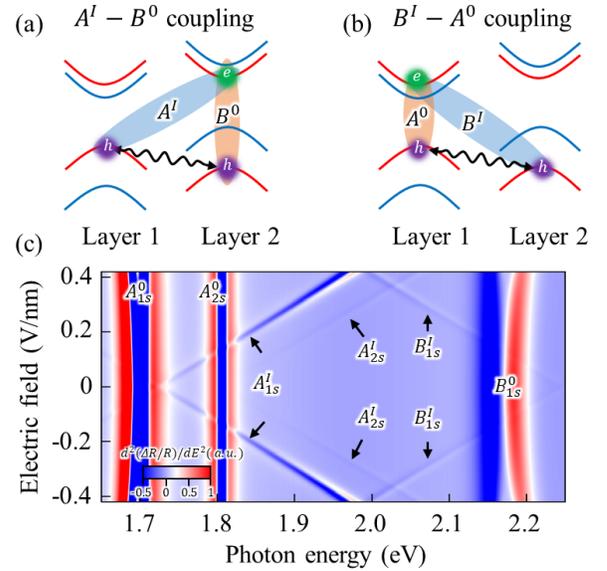

FIG. 3. (a) Schematic illustration of the coupling between the interlayer $A^I$ and intralayer $B^0$ excitons in bilayer WSe$_2$. (b) Schematic for the coupling between the interlayer $B^I$ and intralayer $A^0$ excitons. (c) Simulated field-dependent color maps of the second-derivative reflectance contrast, based on two-level models with constant coupling strengths of W = 60, 11, 60 meV for the $A_{1s}^I - B_{1s}^0$, $A_{2s}^I - B_{1s}^0$, and $B_{1s}^I - A_{1s}^0$ hybridizations, respectively.

To test whether a similar hybridization framework applies here, we construct two-level coupling models to simulate our results (see Supplemental Materials [37] for details). As illustrated in Fig. 3(a), we model the $A_{1s}^I$ and $B_{1s}^0$ excitons as two states with field-dependent energies, with $A_{1s}^I$ being dark and $B_{1s}^0$ being bright. These states can couple because they share the same irreducible representation under the three-fold rotation symmetry of bilayer WSe$_2$. We introduce a constant coupling term $W$ to solve the resulting 2 × 2 Hamiltonian to obtain the hybridized exciton energies and oscillator strengths. In a similar manner, we construct two-level models for the $A_{2s}^I - B_{1s}^0$ and $A_{2s}^0 - B_{1s}^I$ couplings [Fig. 3(a-b)]. Using coupling strengths of W = 60 meV, 11 meV, and 60 meV for the $A_{1s}^I - B_{1s}^0$, $A_{2s}^I - B_{1s}^0$, and $A_{2s}^0 - B_{1s}^I$ pairs, respectively, the simulations successfully reproduce the key spectral features observed in the experiment [Fig. 3(c)]. Notably, the calculated field dependence of the $A_{1s}^I$ oscillator strength closely matches the experimental data shown in Fig. 2(f).

Despite the good agreement with experiment, the required coupling constants raise concern. In particular, the fitted value of W = 60 meV is unrealistically large—nearly two-thirds of the interlayer exciton binding energy (92 meV). As shown later in Fig. 4, DFT calculations yield W = 10 meV at zero electric field, which is six times smaller than the fitted value. With increasing field strength, DFT predicts that W rises to 22 meV at F = 0.4 V/nm due to distortion of the hole wavefunction, and reaches a maximum of 38 meV at F = 0.69 V/nm when the hole becomes equally distributed between the two layers [Fig. 4(a,c)]. These DFT-derived values remain substantially lower than the W = 60 meV extracted from the two-level model. Thus, although the hybridization model reproduces the data, the required parameters are physically implausible. This discrepancy underscores the need for an alternative mechanism to account for the observed interlayer exciton brightening.

*Hole-transfer model*— To uncover the actual mechanism that brightens the interlayer excitons, we perform DFT calculations of the Bloch states in bilayer WSe$_2$ under applied electric fields. We find that the valence-band Bloch states at the K valley are significantly distorted by the electric field, while the conduction-band states at the same valley remain nearly unchanged (see Supplemental Materials [37] for details).

Fig. 4(a) shows the in-plane-averaged hole density of the spin-up valence-band state at the K point under different electric fields. At zero field, the hole is highly localized in one layer (referred to as the majority layer), with only ~2% of the density residing in the other (minority) layer. This strong layer polarization allows us to use the layer indices L1 or L2 to label the excitonic states, distinguishing between intralayer and interlayer excitons based on the electron and hole locations. However, as the electric field increases, the hole wavefunction gradually shifts from the majority to the minority layer. At F = 0.4 V/nm, about the maximum field used in our experiment, the minority-layer hole density increases to ~10% of the total.

This field-driven interlayer hole transfer causes the original interlayer exciton to acquire some intralayer character. While the interlayer component remains optically dark due to spatial separation of the electron and hole, the intralayer component becomes optically active (bright). Importantly, this brightening does not arise from exciton hybridization, as described by the two-level coupling model in Fig. 3, but rather from the intrinsic distortion of the valence-band Bloch states induced by the electric field. Using these distorted valence bands, we computed the resulting exciton eigenstates and their optical spectra. The simulated map in Fig. 4b quantitatively reproduces the key experimental features.

Following the distortion of the valence bands, the resulting excitons—with mixed intralayer and interlayer character—can still hybridize. However, because the Bloch states evolve with the applied electric field, the excitonic coupling strengths also become field-dependent, in contrast to the constant values assumed in simple two-level models. Figure 4c shows the calculated field-dependent coupling strengths for $A^I_{1s} - B^0_{1s}$, $A^I_{2s} - B^0_{1s}$ and $A^0_{2s} - B^I_{1s}$ hybridizations. Among these, the $A^I_{1s} - B^0_{1s}$ coupling term is the strongest, increasing from W = 10 meV at zero field to W = 22 meV at F = 0.4 V/nm, while the other two remain considerably weaker. All of these field-dependent values are substantially smaller than the constant coupling strengths used in the two-level models.

Using these field-dependent couplings and modulated valence bands, we compute the hybridized exciton states (see Supplemental Materials [37] for details). Fig. 4(d) shows the corresponding corrections to the oscillator strength arising from hybridization. The inset compares the total oscillator strength including both hybridization and hole-transfer effects (solid lines) with that obtained by excluding hybridization (dashed lines). For the $A^I_{1s}$ ($A^I_{2s}$) exciton, hybridization contributes only 4% (3%) of the total oscillator strength at zero field, increasing to 15% (19%) at F = 0.4 V/nm as these states shift closer to the $B^0_{1s}$ exciton. For the $B^I_{1s}$ exciton, hybridization remains negligible—0% at zero field and rising to only 3% at F = 0.4 V/nm—due to its larger energy separation from the $A^0_{1s}$ state. These results demonstrate that, within our experimental conditions, the primary driver of interlayer exciton brightening in bilayer WSe$_2$ is electric-field-induced interlayer hole transfer, rather than excitonic hybridization.

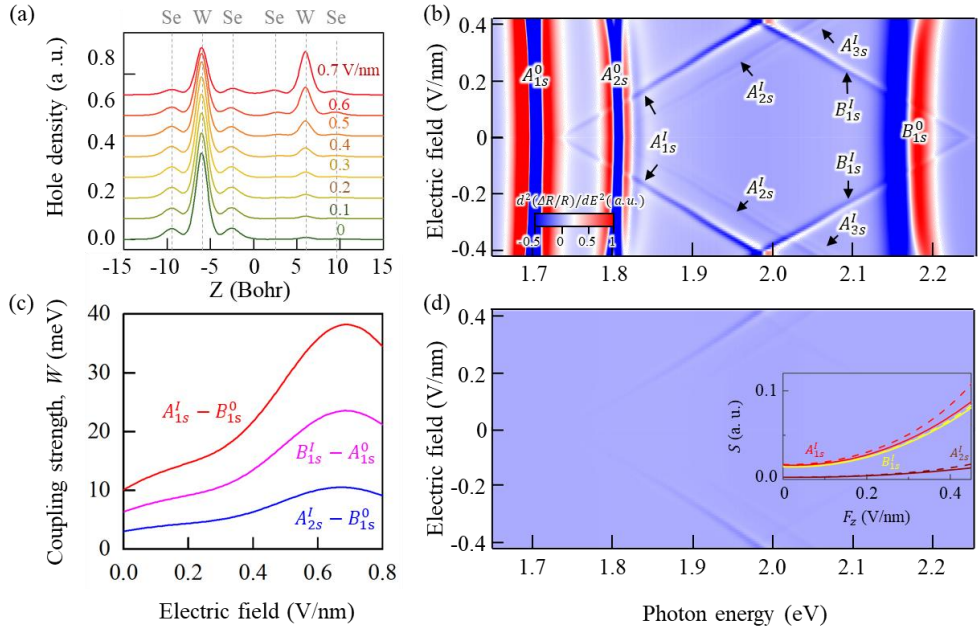

FIG. 4. (a) The DFT-calculated in-plane-averaged hole density of bilayer WSe$_2$ at electric fields F = 0 to 0.7 V/nm (vertically offset clarity). Vertical dashed lines indicate atomic positions. (b) Simulated field-dependent color maps of the second-derivative reflectance contrast, based on field-modified valance-band states incorporating interlayer hole transfer shown in panel (a). (c) Calculated coupling strength for selected exciton pairs as a function of electric field. (d) Corrections to the oscillator strength in panel b arising from exciton hybridization, sharing the same color scale as panel b. The inset compares the total oscillator strength including both hybridization and hole-transfer effects (dashed lines) with that obtained by excluding hybridization (solid lines).

Finally, we remark that field-driven interlayer hole transfer should be a general mechanism for brightening interlayer excitons in TMD materials. Extending our calculations to bilayer MoS$_2$, we find that even at zero field, the valence bands already exhibit substantial hole density in the minority layer, accounting for over 90% of the calculated total oscillator strength of the interlayer A and B excitons. In this regime, hybridization between $A_{1s}^I$ and $B_{1s}^0$ remains weak due to their large energy separation of ~120 meV. Significant hybridization appears only when the interlayer exciton approaches the intralayer exciton in energy, resulting in enhanced brightness and noticeable state repulsion. These findings establish interlayer hole transfer as a robust and general mechanism for brightening interlayer excitons in TMD bilayers and multilayers. This insight is particularly important in regimes where interlayer excitons are well detuned from bright intralayer states and provides a valuable framework for designing optoelectronic devices based on layered TMDs.

*Acknowledgement*s—C.H.L. acknowledges support from the National Science Foundation (NSF) Division of Materials Research CAREER Award No. 1945660. Y.C. Chang acknowledges support from National Science and Technology Council, Taiwan (grant no. NSTC 112-2112-M-001-054-MY2). E.L. acknowledges support from the National Key R&D Program of China (grant no. 2024YFA1410500), the National Natural Science Foundation of China (grant no. 12374456), and the Fundamental Research Funds for the Central Universities. K.W. and T.T. acknowledge support from the JSPS KAKENHI (Grant Numbers 21H05233 and 23H02052), the CREST (JPMJCR24A5), JST and World Premier International Research Center Initiative (WPI), MEXT, Japan. N.M.G. and S.C. were supported by ARO MURI grant no. W911NK- 24-1-0292 and the Army Research Office Electronics Division Award no. W911NF2110260. N. M. G. was also supported by the Presidential Early Career Award for Scientists and Engineers (PECASE) through the Air Force Office of Scientific Research (award no. FA9550-20-1-0097).